\documentclass[11pt,a4paper]{article} 
\usepackage{amsmath,amsfonts,amssymb}  
\usepackage[dvips]{graphicx}

\newcommand{\np}{\nonumber\\}
\newcommand{\pn}{\par\noindent}

\newtheorem{lem}{Lemma}

\title{
Excited states nonlinear integral equations for an integrable  anisotropic
spin 1 chain
}
\author{ J. Suzuki\thanks{e-mail: sjsuzuk@ipc.shizuoka.ac.jp}\\
        \parbox{0.9\textwidth}{
        {\em
        \begin{center}
      Department  of Physics, Faculty of Science\\
   Shizuoka University\\
   Ohya 836, Shizuoka,  Japan
        \end{center}
        }}
       }
\begin{document}
\maketitle
\begin{abstract}
We propose a set of nonlinear integral equations
 to describe on the excited states of
 an integrable the spin 1 chain with anisotropy.
 The scaling dimensions, evaluated numerically in previous studies, are
 recovered analytically by using the equations.
 This result may be relevant to the study on the supersymmetric sine-Gordon model.
  \end{abstract}
%\pacs{75.10.Jm, 05.50+q}

\section{Introduction}

The 1D spin systems have been providing 
problems of both physical and mathematical 
interests. See e.g., \cite{Affleck, Tsvelik}.
Among them, there exists a family of solvable 
models of the Heisenberg's type  with spin-$S$ \cite{Tak,Bab}.
In this report, we are interested in the excited states of 
a member in the family, the spin 1 chain 
with anisotropic interaction.

The recent progress in the study of the integrable system brings forth a powerful 
machinery,   the method of the nonlinear integral equations (NLIE, for short) \cite{KB, KBP}.
The NLIE method has been successfully applied to
the study of the XXZ model ($S=\frac{1}{2}$).
It clarifies the finite size property of the ground state as well as the scaling behavior of
excited states (of corresponding six vertex model \cite{KWZ}).
The application is not restricted to the lattice models: it also provides the detailed descriptions
of the excited states  in the field theoretical models such as the sine-Gordon model\cite{DdV0, DdV1, DdV2,
FMQR, FRT}
and perturbed conformal field theories\cite{BLZ2, BLZ3}.

The study on the higher spin cases has, however, encountered technical difficulties.
This has been resolved in \cite{JSuz}  at least for the ground state.
There the NLIE, which is relevant to the evaluation of the free energy for arbitrary $S$, 
is derived for the isotropic case .  See also \cite{hegedus} for an interesting application
of the result  to the $0(4)$ nonlinear sigma model in the limit $S \rightarrow \infty$ .

In this report, we extend the study to excited states of the anisotropic
 $S=1$ chain . 
Simple assumptions,  suggested by numerical investigations, lead to a set of NLIE
which enables the evaluation of energies for arbitrary system size.
The proposed NLIE has a structure which seems to be a natural extension of the
excited NLIE for the spin $\frac{1}{2}$ case.
We  will  analytically verify the previous observations 
on  some low lying excitations  by numerical methods
\cite{AlcarazMartins, FowlerYuFrahm}.

The result obtained here may be not only relevant in the spin chain problem.
Recently the study on the excited states in supersymmetric sine-Gordon model
attracts much attentions\cite{Dunning, BDPTW}.
It is expected that the proper  discretization of the model is given by the inhomogeneous
 anisotropic spin 1 chain.
 We thus hope that the current study will  shed  some light on the analysis of the supersymmetric sine-Gordon model.

\section{The model and the assumptions}

We are interested in the spin 1 chain with anisotropic interactions.
The hamiltonian contains several
 two-body interactions,
 
\begin{equation}
{\cal H} \negthickspace=\negthickspace\sum_{i=1}^N  
\Bigl (
\sigma_i^{\perp} - (\sigma_i^{\perp})^2 +\cos 2 \gamma (\sigma_i^z -  (\sigma_i^{z})^2)
-(2 \cos \gamma -1) \bigl ( \sigma_i^{\perp} \sigma_i^{z}+ \sigma_i^{z} \sigma_i^{\perp} \bigr )
-4 \sin^2 \gamma  (S_i^z)^2 
\Bigr )
\label{hamil}
\end{equation} 
where  a short-hand notation
$$
\sigma_i =S_i \cdot S_{i+1} =  \sigma_i^{\perp} + \sigma_i^{z} 
$$
is employed.
For simplicity, we impose the periodic boundary condition, $S_{N+1}^a = S_{1}^a , a=x,y,z$.
 The parameter $\gamma$ specifies the anisotropy.
 Throughout this report, we only consider the range
 $\gamma < \frac{\pi}{3}$ and exclude $\gamma$ of the form
 $\gamma = \frac{n \pi}{m}$ with $(m,n)$ co-prime.

 We follow  the strategy in \cite{JSuz}  and start from a integrable 19 vertex model \cite{FZ,SAA}. 
The 19 vertex model is 
obtained from the 6 vertex model by the fusion procedure\cite{SAA, KRS}.
The latter one is associated to the spin $\frac{1}{2}$ XXZ model while the former 
corresponds to the spin 1 chain.
The strategy is to treat the spin 1 model and the spin $\frac{1}{2}$  model simultaneously.
To be more precise, we introduce commuting transfer matrices $T_1(x)$ and $T_2(x)$
acting on  spin 1 quantum space consisting of $N$ sites.  
The  auxiliary space for the former one is given by the
spin $\frac{1}{2}$ space while it is  spin  1 for the latter.

Their explicit eigenvalues read,
\begin{eqnarray}
T_1(z) &=& \phi(z-i \gamma) \frac{Q(z+i \gamma)}{Q(z)}+   \phi(z+i \gamma) \frac{Q(z-i \gamma)}{Q(z)} \nonumber \\
T_2(z) &=& 
 \phi(z-i\frac{\gamma}{2})  \phi(z-i\frac{3\gamma}{2})  \frac{Q(z+ i \frac{3\gamma }{2})}{Q(z-i\frac{\gamma }{2})}  
 + \phi(z-i\frac{\gamma}{2})  \phi(z+i\frac{\gamma}{2})   
 \frac{Q(z+ i \frac{3\gamma }{2}) Q(z-i\frac{3 \gamma }{2})}  {Q(z+ i \frac{\gamma }{2}) Q(z-i\frac{ \gamma }{2})} \np
&   &+ \phi(z+i\frac{\gamma}{2})  \phi(z+i\frac{3\gamma}{2})  \frac{Q(z- i \frac{3\gamma }{2})}{Q(z+i\frac{\gamma }{2})}  
\label{t2expr} \\
\phi(z)&=&\sinh^N z \nonumber
\end{eqnarray}
Note that in place of standard spectral parameter $u$ , we choose  $u=ix$ .
The important function $Q$ is given by the Bethe ansatz roots $z_j, (j=1, \cdots, M)$;
$Q(z) =\prod_{j=1}^{M} \sinh(z-z_j) $.
We will denote three terms of $T_2(z)$ in (\ref{t2expr}) by $\lambda_i(x), i=1,2,3$.

The eigenvalue $E$ of the Hamiltonian  is evaluated
though 
\begin{equation}
E=  \frac{1}{i} \frac{d}{dx} \log T_2(x) |_{x \rightarrow -i\gamma/2}.
\label{energy}
\end{equation} 
Thus once if $Q$ is obtained, the evaluation of  $E$ is  straightforward.
This is equivalent to say that, if all locations of the Bethe ansatz roots are known for excited states, then the
energy is evaluated.
This is, however, a formidable task for  large systems. 
 Apart from  few lower excitations, it is extremely difficult to
find all  locations of Bethe ansatz roots.

The most crucial observation in the NLIE formulation is that this task is avoidable.
With proper choice of auxiliary functions, 
one can bypass dealing with a complete set of  Bethe ansatz roots; one only has to deal with
finite number of complex roots characterizing the excitation\cite{KP, KWZ, DdV2, BLZ2, BLZ3}.
This may break down if very high excitations are of  our interest.
We nevertheless believe that this method will be efficient  
 for the treatment on excited states which are physically important in the thermodynamic limit.
 
 Let us  be more accurate.
 In the ground state, the Bethe ansatz roots are given by the sea of  2 strings.
By a 2 string we mean a pair of Bethe ansatz roots $x_i \pm i y_j$
 with $|y_j -\frac{\gamma}{2}|  \ll 1$.  \pn
 We consider excitations for which only a finite number of roots
deviate from the sea of 2 strings. \pn
  
  In addition, we will make  the following three  assumptions.  \vskip 0.3cm \pn
  {\bf  Assumption 1 }  \\
There should not be  a pair of complex roots $z_1, z_2$ such that $z_1-z_2$ is a multiple of 
$\gamma$.  
 \vskip 0.3cm

For the string type solutions, it is known that the separation of neighboring roots in a string deviates 
slightly from  $\gamma$  for finite system sizes \cite{BVV}.
 Our assumption thus does not contradict
with this pattern. 
It, however,  excludes the complete strings\cite{FM, CK}.
Therefore we should devise another route to deal with the 
case when $q (={\rm e}^{i\gamma})$ is at roots of unity, 
which is beyond the present scope.    \vskip 0.3cm\pn
{\bf  Assumption 2}  \\
The following classification of complex roots , other than 2 strings,  are possible.
\begin{enumerate}
 \item inner roots :  $|\Im z_j| < \frac{\gamma}{2}, j=1, \cdots, M_I$
 \item close roots: $\frac{\gamma}{2} < |\Im z_j| < \frac{3 \gamma}{2},
        j=1, \cdots, M_C$
 \item wide roots:  
$\frac{3\gamma}{2} <|\Im z_j| <\frac{\pi}{2}, j=1,\cdots, M_W$
 \item self conjugate roots :  $|\Im z_j| = \frac{\pi}{2}, \, j=1, \cdots, M_p$.
\end{enumerate}
 \vskip 0.3cm
 
The  above classification  has already been proposed in \cite{AvdeevDorfel} 
which discusses the excitations in the limit $N \rightarrow \infty$.
There a complex conjugate pair of  inner  roots is referred to as a narrow pair
while a pair of  close roots is to a intermediate. 
We adopt a different notation for  similar  roles played by them 
in comparison with the spin $\frac{1}{2}$ case. 
We will sometime denote the locations of inner roots by
 $s^{\pm}_j$,  close roots by $c^{\pm}_j$, 
wide pairs by$w^{\pm}_j$ and self-conjugate roots by $p_j+\frac{\pi}{2} i$.
Here the upper index $+ (-)$ means   its imaginary part being positive (negative).

The zeros of $T_1(x)$ and $T_2(x)$ 
play also important parts. 
The numerical investigation for $N$ up to 8  leads to the  remarkable feature.  \vskip 0.3cm\pn
{\bf  Assumption 3 }  \\
The zeros of   $T_1(x)$ and $T_2(x)$  in the strip $\Im x \in [-\gamma/2,\gamma/2]$ distribute
exactly on the real axis.
We denote  their locations by $\theta^{(1)}_j, (j=1, \cdots, N_1)$ and by $ \theta^{(2)}_{\alpha}, (\alpha=1, \cdots, N_2)$  
 , respectively.
  \vskip 0.3cm
  
\begin{figure}[hbtp]
\centering
{  \includegraphics[width=5cm]{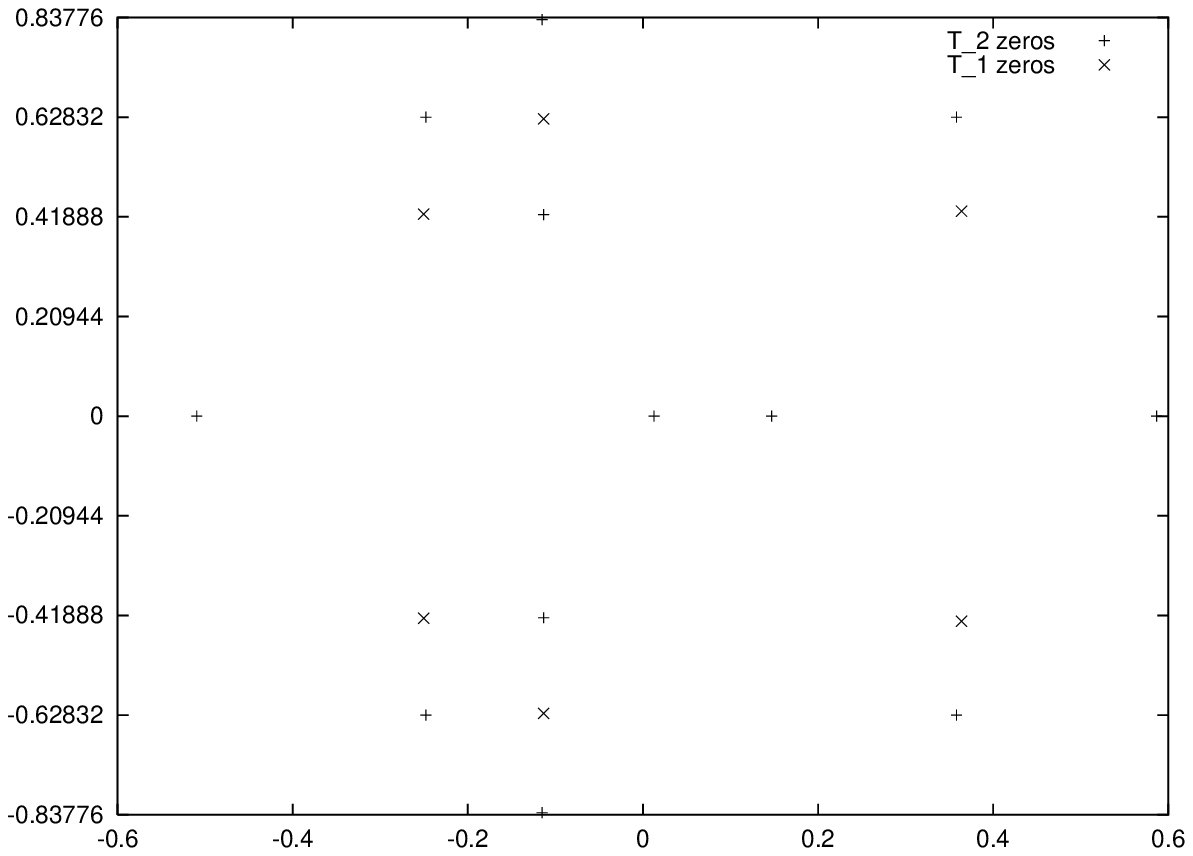} \hspace{1cm}
  \includegraphics[width=5cm]{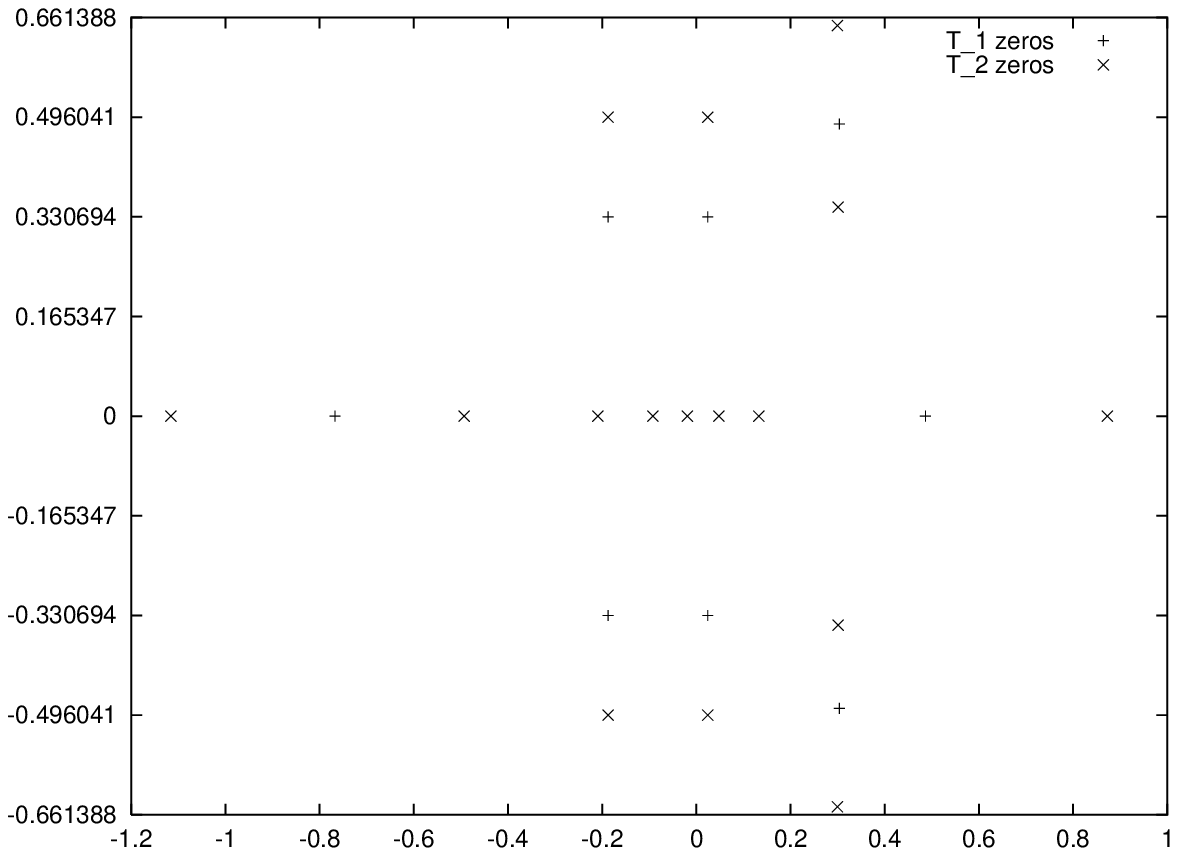} }
\caption{Zeros of $T_1(x)$ and $T_2(x)$, Left:
an excited state in $N=6$ chain $M=6, \gamma=\frac{\pi}{7.5}$
Right: $N=8$ chain $M=7 (S_z=1), \gamma=\frac{\pi}{9.5}$.
}
\label{t1t2zero1}
\end{figure}

 For example, we plot the zeros of $T_1(x)$ and $T_2(x)$ for two cases in fig. \ref{t1t2zero1}.
The state  in the left figure in \ref{t1t2zero1} corresponds to an excited state in 
the singlet sector($M=6$), system size is l $N=6$ chain and with the
  coupling constant $ \gamma=\frac{\pi}{7.5}$.
 For the state in the right figure in  \ref{t1t2zero1}, parameters are chosen $M=7, N=8$ and $\gamma=\frac{\pi}{9.5}$.
 The unit of tics in imaginary direction is normalized to $\frac{\gamma}{2}$.
 These figures thus clearly support the assumption3, which simplifies the derivation of the nonlinear integral equations drastically.

 % \begin{figure}[hbtp]
%\centering
%  \includegraphics[width=8cm]{N6m0_2.eps}
 %\caption{}
%\label{t1t2zero1}
%\end{figure}

 %\begin{figure}[hbtp]
%\centering
%  \includegraphics[width=8cm]{N8m1_5.eps}
 %\caption{}
%\label{t1t2zero2}
%\end{figure}
 
 We supplement the explicit locations of corresponding BAE roots.
 
 \begin{table}[htbp]
\begin{center}
\begin{tabular}{|c|c|}
\hline
\multicolumn{2}{|c|}
{BAE roots for $N=6$, $M=6$, $\gamma=\frac{\pi}{7.5}$}\\
\hline\hline 
   0.65640208387 $+$1.5707963268 i &
 -0.53549516834 $+$1.5707963268i  \\
    -0.11533405190$+$0.20504539087 i &
    -0.11533405190$-$0.20504539087i\\
	   -0.24721889635&
	     0.35698008462\\
  	\hline
\end{tabular}
%\caption{BAE roots for $N=6$, $M=6$, $\gamma=\frac{\pi}{7.5}$.}
%\label{tableroots1}
%\end{center}
%\end{table}
%
%\begin{table}[htbp]
%\begin{center}
\vskip 0.2cm
\begin{tabular}{|c|c|}
\hline 
\multicolumn{2}{|c|}
{BAE roots for $N=8$, $M=7$, $\gamma=\frac{\pi}{9.5}$}\\
\hline\hline
  0.61918637956$+$  1.5707963268i&
-0.10487718381$+$ 1.5707963268i \\
-0.83691028264$+ $   1.5707963268i&
  0.29966102387 $- $0.152149003388i\\
  -0.18745636322  &
  0.024363369664 \\
    0.29966102387$+$ 
    0.15214900339 i &   \\
  	\hline
\end{tabular}
%\caption{BAE roots for $N=8$, $M=7$, $\gamma=\frac{\pi}{9.5}$.}
\label{tableroots1}
\end{center}
\end{table}

 \section{Auxiliary functions and Sum rules} 
 
 In this section, we introduce several auxiliary functions which are crucial for our purpose.\pn
 Firstly we define the most natural auxiliary function $ {\mathfrak a}(z)$,   defined by 
 $$
 {\mathfrak a}(z) = \frac{\lambda_2(z+i\frac{\gamma}{2})} {\lambda_1(z+i\frac{\gamma}{2})}
 =\frac{\lambda_3(z-i\frac{\gamma}{2})} {\lambda_2(z-i\frac{\gamma}{2})}
 $$
 
 In view of  $ {\mathfrak a}(z)$, the Bethe ansatz  equation  can be cast into the form
 $
 {\mathfrak a}(z_j) = -1,\qquad (j=1, \cdots, M).
 $
 One also uses its logarithmic form,  $Z^{(1)}_N(z_j) = 2 \pi I_j$ ,
 where  ${\mathfrak a}(z) = \exp(i Z^{(1)}_N(z))$, 
 which leads to the root density function formulation in the thermodynamic limit.
 The first auxiliary function,  $ {\mathfrak a}(z)$, 
 thus has the deep connection to the Bethe ansatz equation and plays the fundamental
 role  in the study of the spin $\frac{1}{2}$ chain.
 The previous study  \cite{JSuz}
 shows that, unexpectedly,  this is not the case for general values of $S$
 at their ground states.
Instead,  the most crucial functions for the ground state  for $S=1$ are given by

\begin{equation}
\mathfrak{b}_1(x) := \frac{\lambda_1(x)+\lambda_2(x)}{\lambda_3(x)} 
\qquad
\bar{\mathfrak{b}}_1(x) :=
\frac{\lambda_2(x)+\lambda_{3}(x)} {\lambda_{1}(x)}
\label{defb}
\end{equation}
Physically,   $\mathfrak{b}_1(x)$ is related to the  density function 
associated to the centers of 2-strings.

 For a technical reason, we introduce the shifted functions,
 \footnote{ In case of the finite temperature problem, further shifts in $x$ are needed for
analyticity reason, which is not necessary for the finite size problem.}

 $\mathfrak{b}(x)= \mathfrak{b}_1(x+i\epsilon)$
  $\bar{\mathfrak{b}}(x)= \bar{\mathfrak{b}}_1(x-i\epsilon)$
  and capital ones   $\mathfrak{B}(x)= 1+\mathfrak{b}(x)$,
  $\bar{\mathfrak{B}}(x)= 1+\bar{\mathfrak{b}}(x)$.
   The definition obviously concludes,
 \begin{align}
 T_2(x)&= \phi(x+i\frac{\gamma}{2}) \phi(x+i\frac{3\gamma}{2}) 
 \frac{Q(x-\frac{3i \gamma}{2})}{Q(x+\frac{i \gamma}{2})}  \mathfrak{B}(x-i\epsilon)  \label{T2B} \\
 &= \phi(x-i\frac{\gamma}{2}) \phi(x-i\frac{3\gamma}{2}) 
 \frac{Q(x+\frac{3i \gamma}{2})}{Q(x-\frac{i \gamma}{2})}  \bar{\mathfrak{B}}(x+i\epsilon)  \label{T2BB}
 \end{align}
and  $\mathfrak{B}(x)=0$ when $x=\theta^{(2)}_j -i\epsilon$.
  
 In analogy to  ${\mathfrak a}(z) $ and  $Z^{(1)}_N(x)$
we introduce $Z^{(2)}_N(x) := \frac{1}{i} \log \mathfrak{b} (x)$.
In contrast with $Z^{(1)}_N(x), $$Z^{(2)}_N(x)$  is in general a complex-valued function.
We assume that its real part is an almost monotonic increasing function of $x$
and that the imaginary part vanishes when the real part takes integers (half-integers).
One then applies the similar  argument for  the spin $\frac{1}{2}$  \cite{DdV2} to derive 
the following sum rule,

\begin{equation}
N_2= 2(S+M_p) 
-\frac{2}{\pi} \theta ( 3\gamma S+\theta(2 \gamma S   )) +
 2 M_W + M_c 
\label{sumrule2}
\end{equation}
where  $\theta(x):= \pi \lfloor \frac{1}{2}+\frac{x}{ 2 \pi} \rfloor$ and $\lfloor x \rfloor$ specifies the integer
part of $x$.

The validity of this rule is checked against many examples.
The rule is   crucial in the determination of the 
constant term in the NLIE.

We have a remark. The repeated integers are observed for $Z^{(1)}_N(x)$, which
 are attributed to the existence of special holes/roots\cite{DdV2, FMQR}.
Our case studies indicate no symptom of  "special holes/roots" for  $Z^{(2)}_N(x)$.
We therefore dismiss the possibility in this report.
 Even if they exist, only a small modification will be needed in the following argument.

 We need to introduce another pair of auxiliary functions, one of  which coincides with $T_2$
 , apart from the normalization.%, in the case $S=1$.
 \begin{equation}
 y_1(x)  :=\frac{ T_{2}(x) }{\phi(x-i\frac{3\gamma}{2})  \phi(x+i\frac{3\gamma}{2}) } \qquad
Y_1(x) :=1+y_1(x)
\label{defy}
\end{equation}

The essential observation to derive the NLIE for $\ln \mathfrak{b}$ remains almost same 
as the one made in the case of the largest eigenvalue sector of the quantum transfer matrix\cite{JSuz}.

\begin{figure}[hbtp]
\centering
  \includegraphics[width=8cm]{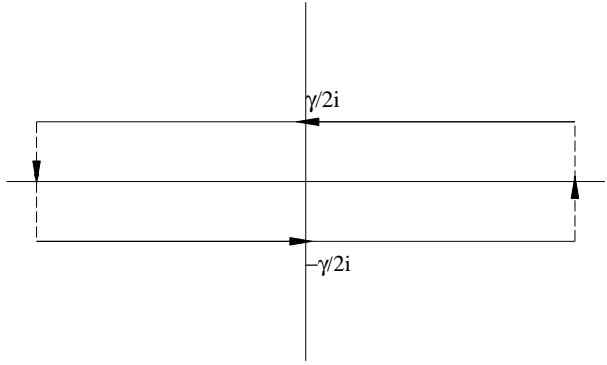}
 \caption{The integration contour ${\cal C}$.
 The straight line in the lower half plane is termed as ${\cal C}_{\ell}$ while
 the one in the  upper plane  is referred to as ${\cal C}_{u}$ after reversing the
 direction.}
\label{piccircleC}
\end{figure}

We introduce renormalized functions, 
\begin{equation}
T^{\vee}_2(x)=
\frac{T_2(x)}{r_2(x) } \qquad 
 r_2(x) := \prod_{\alpha=1}^{N_2}  \tanh \frac{\pi}{2\gamma}(x-\theta^{(2)}_{\alpha})
\label{renormT1T2} 
\end{equation}
and consider  the integral,
$$
\int_{\cal C} \frac{d^2}{d z^2} \log T^{\vee}_2(z) e^{-i k z} dz,
$$
where ${\cal C}$ encircles the strip of  width $\gamma$ (Fig \ref{piccircleC})
in the counterclockwise manner .

Thanks to the renormalization factor,
 $T^{\vee}_2(x)$ contains no zeros or poles inside ${\cal C}$.
The  Cauchy's
theorem thus concludes
\begin{equation}
0= 
\int_{C_{\ell}}  \frac{d^2}{d z^2} \log T^{\vee}_2(z) e^{-i kz} dx
 -\int_{C_u} \frac{d^2}{d z^2} \log T^{\vee}_2(z) e^{-ik z} dx.
 \label{T2cauchy}
\end{equation}
where $C_{\ell},  C_{u}$  means the lower and the upper part of the contour, respectively.

One then substitutes $T^{\vee}_2$ in terms of  $\mathfrak{B}$ for $C_{u}$ and
$T^{\vee}_2$ in terms of  $\bar{\mathfrak{B}}$ for $C_{\ell}$ by utilizing 
(\ref{T2B} ) and  (\ref{T2BB} ).
This provides the various relations among auxiliary functions in the Fourier space.
After straightforward manipulations we find the desired NLIE
\begin{eqnarray}
\ln \mathfrak{b}(x)  &=& 
C_b +
i D_b (x+i\epsilon)   
           \nonumber\\
&  &+ \int_{-\infty}^{\infty} G_s(x-x') \ln \mathfrak{B}(x') dx' 
-\int_{-\infty}^{\infty} G_s(x-x'+2 i \epsilon) \ln \bar{\mathfrak{B}}(x') dx'   \nonumber \\
&  &+ \int_{-\infty}^{\infty} K(x-x'-\frac{\gamma}{2}i +i\epsilon) \ln Y_1(x') dx'    \label{bNLIE2}  \\
\ln y_1 (x)  &= &  D_y (x)  
+ \int_{-\infty}^{\infty}  \bigl( K(x-x'+\frac{\gamma}{2}i -i\epsilon ) \ln \mathfrak{B}(x')+
         K(x-x'-\frac{\gamma}{2}i +i\epsilon ) \ln \bar{\mathfrak{B}}(x')  
 \bigr )  dx'     \nonumber  \\   \label{y1NLIE2}
\end{eqnarray}
where kernel functions read 
\begin{eqnarray*}
G_s(x) &=& \frac{1}{2\pi} \int_{-\infty}^{\infty}  \frac{\sinh(\frac{\pi}{2} - \frac{3\gamma}{2}) k }
                    { 2\cosh (\frac{\gamma}{2} k) \sinh(\frac{\pi}{2} - \gamma) k  }   {\rm e}^{ikx}   \nonumber  dk\\
K(x)  &=&\frac{1}{2\pi}  \int_{-\infty}^{\infty} \frac{1 }
                    { 2\cosh (\frac{\gamma}{2} k)   }   {\rm e}^{ikx}  dk . 
\end{eqnarray*}

It is worth mentioning that  $G_s(x)$ is related to the logarithmic derivative of
the soliton-soliton scattering matrix of the sine-Gordan model, and $K(x)$ is a standard kernel function
in the thermodynamic Bethe ansatz equation for  the RSOS model \cite{BR}.

The constant $C_b$ is found by matching both sides of  NLIE at $x\rightarrow -\infty$
and it reads,
\begin{equation}
C_b = \pi i S - i \theta(\gamma S).
\label{constb}
\end{equation}
We remark that, depending on choice of branches, this value is determined only modulo $2 \pi i$.
This ambiguity can be absorbed into the definitions of branch cut integers.

The driving term in   (\ref{bNLIE2}) consists of three parts:
$
D_b (x) =D^{(b)}_{\rm bulk}(x) +D^{(b)}_{\rm hole}(x)+ D^{(b)}_{\rm roots}(x).
$
The first term is  identical to the one for the ground state, thus we refer it to as
the bulk contribution,

$$
D^{(b)}_{\rm bulk}(x)  =N \chi_K(x)
$$
where  $\chi'_K(x) = 2\pi K(x)$ and  $\chi_K(-\infty)=0$.
The second consists of two pieces, the contributions of holes of $T_1$ and $T_2$.
\begin{equation*}
D^{(b)}_{\rm hole}(x) =
  \sum_j  \chi_K(x-\theta^{(1)}_j )  
  + \sum_{\alpha}  \chi(x-\theta^{(2)}_{\alpha}) 
\end{equation*}
where $ \chi(x)$ is an odd primitive of $ 2\pi G_s(x)$ and   $ \chi(0)=0$.
The third term represents the contributions from complex zeros other than 2 strings,
\begin{eqnarray}
D^{(b)}_{\rm roots}(x) &= & 
-\sum_{\sigma=\pm, j } \chi(x-(c_j^{\sigma} -\sigma \frac{\gamma}{2}i)) +
\sum_{\sigma=\pm, j }  \phi_{\delta}(\frac{x-(w^{\sigma}_j -\sigma \frac{\pi}{2} i)}{\eta}) +
\sum_j   \phi_{\delta}(\frac{x-p_j}{\eta})\\
& &   \label{Droots}  \\
&  & \phi_{\delta}(x) :=\frac{1}{i}  \log -\frac{\sinh(x-i\delta)}{\sinh(x+i\delta)}\nonumber
\end{eqnarray}
and parameters are  given by $\delta=\frac{\pi (\pi-3\gamma)}{2(\pi-2\gamma)}$ and $\eta=1-\frac{2 \gamma}{\pi} $.
The last two summations in (\ref{Droots})  can also be written as 
%$$
%-\sum_j  \bigl (
%     \chi_{II} (x-(w^{+}_j  - i\frac{\gamma}{2}   )) +  \chi_{II} (x-(w^{-}_j  + i\frac{\gamma}{2}   ))
%	        \bigr )
%- \sum_j \chi_{II} (x-p_j -\frac{\pi-\gamma}{2}i )
%$$
$$
-\sum_{\sigma=\pm, j }
    \chi_{II} (x-(w^{\sigma}_j  -\sigma i\frac{\gamma}{2}   )) 
- \sum_j \chi_{II} (x-p_j -\frac{\pi-\gamma}{2}i )
$$
where we adopt a notation; for any function
$f(x)$,  $ f_{II}(x)=f(x)+f(x-i\gamma {\rm sgn }(\Im x))$ \cite{DdV2}.

The driving term for $\ln y_1$ lacks the bulk contribution and 
composed of two terms,    $D_y (x)=D^{(y)}_{\rm hole}+D^{(y)}_{\rm roots}$.
Explicitly,

%\begin{eqnarray*}
%
 %D^{(y)}_{\rm hole}(x)  &=& \sum_{\alpha} \log \tanh \frac{\pi}{2\gamma}(x- \theta^{(2)}_{\alpha}) \\
%
%D^{(y)}_{\rm hole}(x)  &=& i \sum_{\alpha} \chi_K(x- \theta^{(2)}_{\alpha}+\frac{\gamma}{2}i) \\
%
%D^{(y)}_{\rm roots}(x) &=&-i\sum_j \chi_K(x-c^+_j +\gamma i)
%%
%+ i \sum_j \chi_K(x-c^-_j  -\gamma i).
%\end{eqnarray*}

$$
D^{(y)}_{\rm hole}(x) \!= \!i \sum_{\alpha} \chi_K(x- \theta^{(2)}_{\alpha}+\frac{\gamma}{2}i) \quad 
D^{(y)}_{\rm roots}(x) \!=\!-i\sum_{\sigma=\pm, j } \sigma \chi_K(x-c^{\sigma}_j +\sigma \gamma i).
$$

For given locations of  excited zeros and holes,  eqs. (\ref{bNLIE2}) and (\ref {y1NLIE2})
fix the values of auxiliary functions (modulo $exp(2\pi i )$).
Then the  evaluation of 
energy spectra for any  $N$ is immediate by the following expression,
\begin{equation}
E=   e_0  +e_{\rm hole} +e_{\rm roots} +e_{B}.
\label{energyexpression}
\end{equation} 

In the above $e_0$ denotes the bulk ground state energy $e_0= -N(\cot \gamma+\cot 2 \gamma)  $.
The second term   $e_{\rm hole}$
stands for the excitation energy for a hole,
$$
e_{\rm hole}= \sum_j  e (\theta^{(2)}_j ) , \qquad 
 e(x):=
\frac{\pi}{\gamma} \frac{1}{\cosh \frac{\pi}{\gamma} x}.
$$
Among the contributions from complex excitations,  the one from the close roots
appears here explicitly,
$$
e_{\rm roots} =-\sum_{\sigma=\pm, j}
 e( c^{\sigma}_j - i  \sigma \frac{\gamma}{2})  .
$$
This  phenomenon is also observed for the spin $\frac{1}{2}$ case \cite{DdV2}
for $\gamma \le \frac{\pi}{2}$.

The last term contains implicit contributions from all excitations, and it is given by
$$
e_{B} =\frac{1}{i} \int K'(x-x'-i\epsilon+i\frac{\gamma}{2}) |_{x \rightarrow -i\gamma/2}    \ln \mathfrak{B}(x') dx'+
\frac{1}{i} \int K'(x-x'+i\epsilon-i\frac{\gamma}{2})|_{x \rightarrow -i\gamma/2} \ln \bar{\mathfrak{B}}(x') dx' .
$$
The locations of complex roots and holes are, however, not given a priori.   
Therefore
an improvement of the set of NLIE is  necessary
 so as to make the evaluation of these locations possible.
To resolve this, we need  the  NLIE for  $\ln \mathfrak{a}$ 
with a general complex argument $z$.

The derivation of the NLIE can be done in two ways.  
One  starts from the derivation 
for real $z$, then apply the analytic continuation argument in \cite{DdV2}
to derive the equation valid for $y$ with larger complex parts.
The procedure is straightforward but for a small technical difficultly
which does not show up in the spin $\frac{1}{2}$ problem.
Alternatively,  one can  start directly from $y$ with larger complex part, apply 
carefully the following simple lemma \ref{lemmabasic}
to reach desired  NLIE. 
\begin{lem}\label{lemmabasic}
We define, for a smooth function $f(z)$,
its "Fourier transformation"
by
\begin{equation}
f_y[k] = \frac{1}{2\pi} 
 \int_{-\infty}^{\infty} f(x+ i y) {\rm e}^{-ikx} dx.
%\label{defFT}
\end{equation}
with $z=x+i y$.

Suppose $f(z)$ has a pole at $z=z_0$ with the residue $r$ and analytic elsewhere.
For $\Im z_0 >0$  then
$$
f_y[k]=
\begin{cases}
  f[k] {\rm e}^{-k y}&  0 \le y < \Im z_0 \\
 (f[k]- ir {\rm e}^{-ik z_0} ) {\rm e}^{-k y}&  y>\Im z_0 \\  
\end{cases}
$$
Similarly for  $\Im z_0 <0$
$$
f_y[k]=
\begin{cases}
  f[k] {\rm e}^{-k y}&   \Im z_0  \le y < 0 \\
 (f[k]+ ir {\rm e}^{-ik z_0} ) {\rm e}^{-k y}&  y<\Im z_0 \\  
\end{cases}.
$$
\end{lem}
We have derived the following equations by these two manners and  checked that
they lead to identical 
results.

A  moment consideration convinces us that we need to treat the equations, at least 
by three separate regimes for positive imaginary values of $z$.

For the simplest case, $\Im z \in [0, \frac{\gamma}{2})$ , the NLIE reads

\begin{equation}
\ln \mathfrak{a}(z) =  C_a + i D_a(z)  - \int_{-\infty}^{\infty} K(z-x'-i\epsilon) \ln \mathfrak{B} dx'
                            + \int_{-\infty}^{\infty} K(z-x'+i\epsilon) \ln \bar{\mathfrak{B}} dx'   \label{aNLIE1}
\end{equation}
where the constant and drive term are given by
\begin{eqnarray*}
 C_a &= &  2 i\theta(2\gamma S) - i \theta(3\gamma S+\theta(2\gamma S))   \nonumber \\
  D_a(z) &= &    \sum_{\sigma=\pm, j} \chi_K(z-(c_j^{\sigma} -\sigma \frac{\gamma}{2}i)) -
  \sum_{\alpha} \chi_K(z-\theta^{(2)}_{\alpha}) .
\end{eqnarray*}

The equation ceases to be valid as  $\Im z $ crosses  $\frac{\gamma}{2}$ due to the singularities 
existing in the kernel functions of the above equation.
Taking account of the modification due to them, we find 
the equation for $ \Im z  \in   (\frac{\gamma}{2}, \frac{3\gamma}{2}  ) $ 

\begin{eqnarray}
\ln \mathfrak{a}(z) \! &-& \!  \ln (1+\frac{1}{\mathfrak{a}(z -i\gamma)}) 
\!=\! -C_b - i D_b(z-i\frac{\gamma}{2})  \nonumber \\
\! & -& \! \int_{-\infty}^{\infty} G_s(z-i\frac{\gamma}{2} -x'-i\epsilon) \ln \mathfrak{B} dx'
                            + \int_{-\infty}^{\infty} G_s(z-i\frac{\gamma}{2} -x'+i\epsilon) \ln \bar{\mathfrak{B}} dx'   \nonumber\\
\! &+&	\!\int_{-\infty}^{\infty} K(z-x') \ln (1+y_1(x')) dx'   				
\label{aNLIE2}
\end{eqnarray}
The simultaneous evaluation of  $\mathfrak{a}(z)$ at different strips
is thus necessary within this framework.
When $z=i\gamma$, some spurious singularities appear in the rhs, which causes some numerical problems.
We hope to report on the remedy in a separate communication.

The last case, $\Im z \in (\frac{3\gamma}{2}, \frac{\pi}{2})$, the equation takes again a simpler form,

\begin{equation}
\ln \mathfrak{a}(z) = i \widetilde{D_a}(z)  - \int_{-\infty}^{\infty} G_s^{II} (z-i\frac{\gamma}{2} -x'-i\epsilon) \ln \mathfrak{B} dx'
                            + \int_{-\infty}^{\infty} G_s^{II} (z-i\frac{\gamma}{2} -x'+i\epsilon)  \ln \bar{\mathfrak{B}} dx'  
							\label{aNLIE3}
\end{equation}
Note that integration constant is null modulo $2\pi i$ and the drive term reads,
\begin{eqnarray*}
\widetilde{D_a}(z)  & =&  
\sum_{\alpha} \phi_{\delta}(\frac{z-i\frac{\pi}{2}-\theta^{(2)}_{\alpha}}{\eta}) 
- \sum_{\sigma=\pm, j } \phi_{\delta}(\frac{z-i\frac{\pi}{2}-( c^{\sigma}_j -\sigma \frac{\gamma}{2}i)}{\eta})    \\
& &-\sum_{\sigma=\pm, j } \phi^{II}_{\delta}(\frac{z-i\frac{\gamma}{2}-( w^{\sigma}_j -\sigma \frac{\pi}{2}i)}{\eta})
-\sum\phi^{II}_{\delta}(\frac{z-i\frac{\gamma}{2}-p_j}{\eta})
\end{eqnarray*}

Although these three NLIE for $\ln {\mathfrak a}$ take rather involved forms, their meaning  is simple:
once $\ln \mathfrak{B}, \ln \bar{\mathfrak{B}}$ and $\ln Y$ are known {\itshape on the real axis},
the NLIE yield the  evaluation of  $\ln {\mathfrak a}$  {\itshape at arbitrary} $z$.
Again, we have checked the validity of these NLIE at various points in the complex plane.

As noted above, the locations of complex roots and holes are determined by "quantization" conditions
\begin{align}
\ln \mathfrak{a} (\theta^{(1)}_j )  &= (2 I^{(a)}_j+1) \pi i, &
\ln \mathfrak{b} (\theta^{(2)}_{\alpha} - i\epsilon) & = (2 I^{(b)}_{\alpha}+1) \pi i \nonumber \\
\ln \mathfrak{a} (c^{\pm}_j )  &= (2 I^{(c^{\pm} )}_j+1) \pi i, &
\ln \mathfrak{a} (w^{\pm}_j )  &= (2 I^{(w^{\pm}) }_j+1) \pi i,  \nonumber \\
\ln \mathfrak{a} (p_j +\frac{\pi}{2}i )  &= (2 I^{(p) }_j+1) \pi  i.
\end{align}

To be precise  the NLIE still leaves  $2\pi i$ ambiguity as  remarked in the above .
 Thus one has to be
careful in the choice of branch cuts.  We will present several examples of proper
choices in the next section.

The set of properly chosen integers  then fixes the locations of complex roots and holes, 
thereby the energy spectra. 
 In this sense, the NLIE characterizes the finite size
spectra completely.

\section{Low lying Excitations in the thermodynamic limit}

For an illustration, we consider the energy levels of some low lying excitations in the thermodynamic limit.
We consider the weak anisotropy $\gamma \rightarrow 0$ case. 
As shown in \cite{KBP, KP, DdV2}, the scaling behavior of these levels can be evaluated without solving the NLIE explicitly.

The contribution to the excited spectra mainly comes from the left and right extremes of
roots distribution.
 In the limit, $N \rightarrow \infty$, these two contributions are almost decoupled.
 
 The energy scales as
$$
E \sim N \epsilon_0 + \frac{1}{N} \triangle E
$$
where $\epsilon_0$ denotes the bulk ground state energy.
The  scaling energy $\triangle E$  is given by the sum of left
and right contributions,  $\triangle E = \triangle E_+ + \triangle E_-  $.
We  conveniently introduce scaling functions,

\begin{align*}
a_{\pm}(x) &= \mathfrak{a }( \pm \frac{\gamma}{\pi}(x+\ln N) ) , & 
A_{\pm}(x) &=1+a_{\pm}(x)   \\
b_{\pm}(x) &= \mathfrak{b} ( \pm \frac{\gamma}{\pi}(x+\ln N))  , & 
B_{\pm}(x)&=1+b_{\pm}(x) \\
y_{\pm}(x) &= y_1 ( \pm \frac{\gamma}{\pi}(x+\ln N)) ,& 
Y_{\pm}(x)&=1+y_{\pm}(x) \\
\end{align*}

Similarly, 
to holes near the left (right) extremes, we associate the new locations
$\theta^+$($\theta^-$), 
\begin{equation}
\theta^{\pm}_j =\pm \frac{\pi}{\gamma}\theta^{(2)}_j -\log N
\label{rescalehole}
\end{equation}

Then the left/right  contributions are explicitly written as 

\begin{equation}
 \triangle E_{\pm} = \frac{2\pi}{\gamma} 
\sum_j {\rm e}^{-\theta^{\pm}_j}  
\pm \frac{1}{i\gamma} \int {\rm e}^{-(x \pm i \epsilon')} \log B_{\pm}  dx
\mp \frac{1}{i\gamma} \int {\rm e}^{-(x \mp i \epsilon')} \log {\bar B}_{\pm}  dx 
   \label{scalingE}
\end{equation}
where  $\epsilon'=\frac{\pi}{\gamma} \epsilon $.
In the following discussion,  it is irrelevant thus we set $\epsilon' \rightarrow 0$ 
for notational simplicity.

The NLIE also can be transformed into scaling forms.
We prepare associated kernel functions and drive functions,
\begin{align*}
K^{\gamma}(x) &=\frac{\gamma} {\pi} K(\frac{\gamma}{\pi} x) ,& 
G^{\gamma}(x) &=\frac{\gamma}{\pi}  G_s(\frac{\gamma} {\pi}x)  \\
\chi^{\gamma}(x)&:= \chi(\frac{\gamma}{\pi}x),&
  \chi^{\gamma}_K(x)&:= \chi_K(\frac{\gamma}{\pi}x)
\end{align*}

Then the set of resultant NLIE takes the form,

\begin{align*}
\ln b_{\pm}(x) =&
\mp 2i {\rm e}^{-x}+ C^{\pm}_b +i D^{\pm}_b(x) + \int G^{\gamma}(x-x') \ln B_{\pm}(x') dx' -
 \int G^{\gamma}(x-x' ) \ln {\bar B}_{\pm}(x') dx'    \nonumber \\
& +\int K^{\gamma}(x-x' \mp \frac{\pi}{2} i ) \ln Y_{\pm}(x') dx'   \\
\ln y_{\pm}(x) =&
 C^{\pm}_y +D^{\pm}_y(x) + \int K^{\gamma}(x-x' \pm i\frac{\pi}{2} ) \ln B_{\pm}(x') dx' \\
 &+ \int K^{\gamma}(x-x' \mp i\frac{\pi}{2} ) \ln {\bar B}_{\pm}(x') dx' .
\end{align*}

Those scaling NLIE for $\ln a_{\pm}$ have similar  but  more involved forms.
We only write the case $\Im x \in [0,\frac{\gamma}{2})$,
\begin{equation*}
\ln a_{\pm}(x) = C_a^{\pm} +i D_a^{\pm} -\int  K^{\gamma} (x-x') \log B_{\pm}(x') dx' +
\int  K^{\gamma} (x-x') \log \bar{B}_{\pm}(x') dx' 
\end{equation*}

The constants and $D$ functions in the above depend on the choice of specific state
of interest and choice of branch cut integers.
Below we consider simple examples  for which the results of numerical studies are available. \pn
%
%For these cases, we do not have deal with the NLIE for $\ln a_{\pm}$ .\pn
%
Example 1: \pn
We consider the lowest excitation in the sector $S_z=1$.
It is shown numerically that the $N-2$ zeros of $Q(x)$ form the sea of the two strings
and the last zero is located at the origin\cite{AlcarazMartins, FowlerYuFrahm}.  
In the strip $\Im x \in [-\frac{\gamma}{2}, \frac{\gamma}{2}]$,
$T_1$ possesses no zeros while two zeros  of $T_2$ 
are located on the real axis.  We denote their scaled locations  (\ref{rescalehole}) by
 $ \theta^{\pm}$.

The drive terms in this case are found to be,
\begin{align*}
C^{\pm} _b &=\mp(\pi-\delta) i,&  D^{\pm}_b &= \pm \chi^{\gamma}(x-\theta^{\pm}) \\
C^{\pm}_ y &= \pm \pi i,&   D^{\pm}_y &= \pm i \chi^{\gamma}_K(x-\theta^{\pm}\pm \frac{\pi}{2}i) .
\end{align*}

We conveniently choose the quantization conditions,
$\log b_{\pm}(\theta^{\pm})=\mp (2I^{\pm}+1) \pi$.
The locations of holes then satisfy
\begin{eqnarray}
{\rm e}^{-\theta^{\pm}}&=& \mp \frac{1}{2i} 
\Bigl (
-(2 \pi I^{\pm}  +\delta) i 
  -\int G^{\gamma}(\theta^{\pm}-x) 
            \ln B_{\pm}(x) dx \nonumber\\
  &  & \phantom{aaa} + \int G^{\gamma}(\theta^{\pm}-x ) 
            \ln {\bar B}_{\pm}(x) dx 
        -\int K^{\gamma}(\theta^{\pm}-x \mp \frac{\pi}{2} i   )
            \ln Y_{\pm}(x) dx 
   \Bigr )
   \label{thetaloc1}
\end{eqnarray}
 Our choice of $C^{\pm}_b$ and of the branch cut integers leads to $I^{\pm} \ge 0$ in the present case.
By substituting  (\ref {thetaloc1}) into  (\ref{scalingE}), we present 
  $\triangle E_{\pm} $  only  in terms of auxiliary functions,
\begin{eqnarray}
\triangle E_{\pm}&=&\frac{1}{i\gamma}
\Bigl(
  \pm \int {\rm e}^{-x } \log B_{\pm}  dx
   \mp \int {\rm e}^{-x} \log {\bar B}_{\pm}  dx 
 +(2 \pi^2 I^{\pm} +\delta \pi)i               \nonumber \\
&  &  \phantom{aaa} \pm  \pi    \int G^{\gamma}(\theta^{\pm}-x)   \ln B_{\pm}(x) dx
         \mp   \pi \int G^{\gamma}(\theta^{\pm}-x) 
            \ln {\bar B}_{\pm}(x) dx   \nonumber \\
 & &    \phantom{aaa}    \pm \int K^{\gamma}(\theta^{\pm}-x \mp \frac{\pi}{2} i  )
            \ln Y_{\pm}(x) dx
      \Bigr )
\end{eqnarray}

The standard dilogarithm trick leads to the explicit values of
desired integrals,
\begin{eqnarray}
& &4i 
\Bigl(
\pm \int {\rm e}^{-x} \log B_{\pm}  dx
   \mp \int {\rm e}^{-x} \log {\bar B}_{\pm}  dx   \nonumber \\
&  &\phantom{aa} \pm  \pi    \int G^{\gamma}(\theta^{\pm}-x)   \ln B_{\pm}(x) dx
         \mp   \pi \int G^{\gamma}(\theta^{\pm}-x) 
            \ln {\bar B}_{\pm}(x) dx       \nonumber  \\
 & &  \phantom{aa}    \pm \pi  \int K^{\gamma}(\theta^{\pm}-x \mp \frac{\pi}{2} i   )
            \ln Y_{\pm}(x) dx			\Bigr )     \nonumber  \\
& & \phantom{aa}  
 \pm (\pi-2\delta)i  \bigl( \log B_{\pm} (\infty) -\log {\bar B}_{\pm} (\infty)  \bigr)
\mp i \pi \log Y_{\pm}(\infty)      \nonumber  \\
			& & =\int (\log b_{\pm})' \log B_{\pm} dx -  \int \log b_{\pm} ( \log B_{\pm})' dx +
           \int (\log {\bar b}_{\pm})' \log {\bar B}_{\pm} dx -  \int \log {\bar b}_{\pm} ( \log {\bar B}_{\pm})' dx   
		         \nonumber  \\
	&  &  \phantom{aa}  +   \int (\log y_{\pm})' \log Y_{\pm} dx -  \int (\log y_{\pm}\mp \pi i) ( \log {Y}_{\pm})' dx 
	\label{integ1}
\end{eqnarray}

where we have used   $\chi^{\gamma}_K(\infty)=\pi$
It is worth mentioning the asymptotic values
\begin{align*}
b_{\pm} (-\infty) &=0 &   { \bar b}_{\pm} (-\infty) &=0  \\
y_{\pm} (-\infty) &=-1 &  &  \\
b_{\pm} (\infty) &=\frac{1+{\rm e}^{\mp 2i \gamma}  }{ {\rm e}^{\pm 2i \gamma}  }   & 
  { \bar b} _{\pm} (\infty) &=\frac{1+{\rm e}^{\pm 2i \gamma}  }{ {\rm e}^{\mp 2i \gamma}  } \\
y_{\pm} (\infty) &=1+2 \cos 2 \gamma .
\end{align*}
Using these values and 
by the change of integration variables,  we find that the rhs of  (\ref{integ1}) is given by
the dilogarithm functions,
\begin{equation}
2 L_{+}(b_{\pm}(\infty))+  2 L_{+}(b_{\pm}(\infty))+ 2L_+(y_{\pm}(\infty))-2L_+(1)+2 L(1) +2 L_+(1)\mp \pi i  \log Y_{\pm}(\infty)
\label{dilogsum1}
\end{equation}
Explicit definitions are as follows.
\begin{eqnarray*}
L_+(x) &:=& \frac{1}{2}\int_0^x 
  ( \frac{ \log(1+y)}{y}-\frac{\log y}{1+y} ) dy,\\
L(x) &:=& -\frac{1}{2} \int_0^x  (\frac{\log(1-y)}{y}+\frac{\log y}{1-y}) dy
\end{eqnarray*}

The scaling energy  $\triangle E$ is divided into two parts $\triangle E= \triangle E_1+ \triangle E_2$.
The former brings the central charge $\triangle E_1=\frac{\pi v}{6}c$ 
while the second is related to the scaling dimension,
$\triangle E_2=2 \pi v X$.  
The sound velocity is readily evaluated $v=\frac{\pi}{\gamma}$. 

It is easily shown that the sum of the first four terms  in (\ref{dilogsum1})
remains constant for small $\gamma$.
Thus the dilogarithm formula, utilized in the study of the ground state \cite{JSuz}
in the rational case ($\gamma \rightarrow$ 0), is also applicable. 
We identify these  terms  as the contribution to  $\triangle E_1=\frac{\pi v}{6}c$.
This lead to
the correct central charge $\frac{3}{2}$.

Evaluating the remaining contributions, we find,
$$
\triangle E_2=\frac{2 \pi^2}{\gamma}( I_+ + I_- +X_c+\frac{1}{8})
$$
where $X_c=\frac{\pi-2 \gamma}{4 \pi}$.
 Therefore, by choosing the
minimal value $I_+=I_-=0$, the above calculation recovers the desired scaling dimension,
$X= X_c+\frac{1}{8}$  \cite {AlcarazMartins, FowlerYuFrahm}.

\pn
Example 2: \pn
Let us consider another excitation in the sector $S_z=1$.
The $ N-2 $ zeros of Q function again form the sea of the two strings
while the last zero is located at $\frac{\pi}{2}i$.  
In this case, $T_1(x)$ possesses 2 zeros on the real axis
while $T_2(x)$ does 4 zeros.
We denote corresponding scaling locations by $\xi^{\pm}$  for zeros of $T_1(x)$ 
and
$\theta^{\pm}_j, (j=1,2)$ for those of $T_2(x)$.
In this case it is convenient to introduce $ \chi^{+}_K (x) := \chi^{\gamma}_K(x) -\pi$ so that
$ \chi^{+}_K(\infty)=0$.

Then the drive terms are then given by,
\begin{align*}
C^{\pm}_b &=\mp \pi  i&    D^{\pm}_b&= \pm  \sum_j \chi^{\gamma}(x-\theta^{\pm}_j )  \pm \chi^{+}_K(x-\xi^{\pm}) \\
C^{\pm}_y&= 0&   D^{\pm}_y &= \pm i \sum_j \chi^{+}_K(x-\theta^{\pm}_j \pm \frac{\pi}{2}i) \\
C^{\pm}_a &= 0&    D^{\pm}_a &= \mp  \sum_j \chi^{+}_K(x-\theta^{\pm}_j ) \\
\end{align*}

The quantization conditions read 
\begin{equation*}
\log b^{\pm}(\theta^{\pm}_j) = \mp (2 I^{\pm}_j +1) \pi
\qquad 
\log a^{\pm}(\xi^{\pm}) = \pm (2 J^{\pm} +1) \pi
\end{equation*}

One easily verifies  $I^{\pm}_j,   J^{\pm} \ge 0$.

The quantization condition for $\log b^{\pm}$ leads to an expression analogous to (\ref {thetaloc1}).
This time, 
however, the rhs contains $\frac{1}{2} \sum_j  \chi^+_K (\theta^{\pm}_j -\xi^{\pm})$.
The quantization  condition for $\log a^{\pm}$  enables us to represent this by integrals,

\begin{equation}
 \sum_j \chi^+_K (\theta^{\pm}_j -\xi^{\pm}) =-\pi  \pm \frac{1}{i} \int K^{\gamma}(\xi^{\pm}-x) \log B_{\pm}(x) dx 
 \mp  \frac{1}{i} \int K^{\gamma}(\xi^{\pm}-x) \log \bar{B}_{\pm}(x) dx 
\end{equation} 
We used a simple relation $\chi^+_K (x)+\chi^+_K (-x)=-\pi$ and  put  $J^{\pm}=0$.

We then proceed as example 1 and obtain the same $\triangle E_1$ and
$$
\triangle E_2= \frac{2 \pi^2}{\gamma} 
\bigl (  
   1+ \frac{\gamma}{\pi (\pi-2 \gamma)}
   \bigr )
= \frac{2 \pi^2}{\gamma} 
\bigl (  
   X_c+\frac{1}{16 X_c} +\frac{1}{2}
   \bigr )
$$
where we choose $ I^{\pm}_j = j-1 $.
This again coincides with the numerical result \cite {AlcarazMartins}.

We also analyzed several examples in the different spin sectors, and checked that
the results all recovered the desired values. 
We, however omit further discussions for brevity.

\section{Discussion and Summary}

In this report, we derive a set of NLIE
which characterizes excited state spectra of the spin 1 chain with anisotropic interactions.
The equations are tested numerically against exact data.  
Some desired scaling dimensions are derived analytically for some low excitations.

Finally, we comment on an implication of the result obtained here.
Through the light cone approach\cite{DdVinh1, DdVinh2}, 
the inhomogeneous version of the spin 1 chain, or the 19 vertex model will be the proper candidate for
the discrete analogue of the supersymmetric sine-Gordon model .
The latter's  excited spectra have been attracted attentions recently\cite{BDPTW} .
A proper deformation of the NLIE obtained in this paper may be useful in such investigations.
Assume  that the several conjectures on
 the analytic properties in the  spin chain problem are also valid for
supersymmetric sine-Gordon model. 
Then it is readily shown that  we reach the almost same nonlinear integral equations.
The "bulk" driving term of  $\log {\mathfrak b}$   should be then replaced by
$m L \sinh \frac{\pi}{\gamma} x$ where $m L$ is related to the inhomogeneity $\Lambda$
by   $m L= 2N \exp(-\frac{\pi}{\gamma}\Lambda)$.
We however avoid drawing a conclusion in haste as
it requires careful analytic and numerical checks; the consistency to the $S-$matrix picture in
\cite{CAhn}, for instance.
We hope to report this in a near future,  together with complete discussion 
on the conformal limit. 

\vskip 0.6cm
\noindent{\bf Acknowledgements} 
\vskip 0.3cm
\pn
The author would like to thank  C. Ahn and F. Ravanini for discussions, 
C.Dunning for the critical reading of the manuscript, 
 P.E. Dorey and J.M. Martins  for their interest.
The work of JS has been supported by a Grant-in-Aid
for Scientific Research from the Ministry of Education, Culture,
Sports and Technology of Japan, no.~14540376.

\end{document}